\documentclass[sigconf,nonacm]{acmart}

\usepackage[T1]{fontenc}
\usepackage[utf8]{inputenc}
\usepackage{amsmath}
\usepackage{booktabs}
\usepackage{tabularx}
\usepackage{array}
\usepackage{graphicx}
\usepackage{url}

\renewcommand\footnotetextcopyrightpermission[1]{}
\newcolumntype{Y}{>{\centering\arraybackslash}X}
\newcolumntype{L}{>{\raggedright\arraybackslash}X}

\title{Accurate Portraits of Scientific Resources and Knowledge Service Components}

\author{Yue Wang}
\affiliation{%
  \institution{Beijing Key Laboratory of Intelligent Telecommunication Software and Multimedia, School of Computer Science (National Pilot School of Software Engineering), Beijing University of Posts and Telecommunications}
  \city{Beijing}
  \country{China}}

\author{Zhe Xue}
\authornote{Corresponding author.}
\affiliation{%
  \institution{Beijing Key Laboratory of Intelligent Telecommunication Software and Multimedia, School of Computer Science (National Pilot School of Software Engineering), Beijing University of Posts and Telecommunications}
  \city{Beijing}
  \country{China}}

\author{Ang Li}
\affiliation{%
  \institution{Beijing Key Laboratory of Intelligent Telecommunication Software and Multimedia, School of Computer Science (National Pilot School of Software Engineering), Beijing University of Posts and Telecommunications}
  \city{Beijing}
  \country{China}}

\begin{abstract}
With the advent of the cloud computing era, the cost of creating, capturing, and managing information has gradually decreased. The amount of data on the Internet is showing explosive growth, and more scientific and technological resources are being uploaded to the network. Different from news and social media data, scientific and technological resources are mainly composed of academic-style resources or entities, such as papers, patents, authors, and research institutions. There is a rich relationship network between these resources, from which a large amount of cutting-edge scientific and technological information can be mined. Existing scientific resource management and classification standards are difficult to completely cover all entities and associations, and they cannot accurately extract the important information contained in scientific and technological resources. Therefore, how to construct a complete and accurate representation of scientific and technological resources from structured and unstructured reports and texts, and how to tap the potential value of scientific and technological resources, are urgent problems. A feasible solution is to construct accurate portraits of scientific and technological resources by combining knowledge graph technology, text representation learning, entity extraction, and knowledge service components.
\end{abstract}


\begin{document}
\maketitle

\section{Introduction}
With the advent of the cloud computing era, technologies and studies related to big data have received increasing attention. The cost of creating, capturing, and managing information has gradually decreased, while the amount of data on the Internet has grown explosively. More and more scientific and technological resources are uploaded to the network. Different from common Internet news and social media data, the main body of scientific and technological resources is composed of academic-style resources or entities, such as papers, patents, authors, and research institutions. These resources form networks of associations, from which a large amount of frontier scientific and technological information can be mined.

Scientific and technological resources involve various entities related to scientific research, such as papers, patents, scholars, institutions, and publishing units. The relationships between these entities form a massive and heterogeneous network of scientific and technological resources \cite{shi2019deepcollaborative,kou2016socialnetworksearch}. Existing retrieval websites can provide search and query services for researchers, but these services mainly satisfy basic retrieval needs and do not sufficiently explore the potential value of scientific and technological resources \cite{kou2016socialnetworksearch}. Knowledge graph technology can integrate entity extraction, relationship mining, information processing, knowledge measurement, and visualization, thereby supporting accurate portraits of scientific resources.

Recent studies further show that scholar portraits are an important part of scientific resource portraits. Multi-view scholar clustering with dynamic interest tracking can model the evolving interests of scholars, which is useful for author portraits, academic community analysis, and researcher-oriented knowledge services \cite{li2023mvsc}. Broader innovation studies also indicate that creative personality, education, entrepreneurial identity, and innovation behavior can provide contextual signals for human-centered academic and innovation portraits \cite{zhou2020creativeentrepreneur}.

By extracting and analyzing scientific resource entities and entity relationship networks, a knowledge graph of a related field can be constructed. By analyzing the relationships between entities in a subject domain, users can quickly understand major research results and important researchers in each subject area. By accurately clustering the topics of resources, the accuracy of query and knowledge services can be improved. In addition, modularity-based community detection can help discover scientific communities in citation and collaboration networks \cite{yang2016modularity}. Heterogeneous graph attention networks and graph neural networks with incomplete features and structures provide useful references for modeling sparse and heterogeneous scientific resource graphs \cite{hu2019hgat,huo2023t2gnn}. Self-supervised reciprocally contrastive learning on heterogeneous graphs can further strengthen representation learning when label information is limited \cite{jin2022reciprocalcontrastive}. These studies motivate the construction of accurate scientific resource portraits and knowledge service components.

\section{Acquisition and Feature Representation of Scientific Resource Texts}
Compared with traditional Internet data, scientific and technological resources exhibit more complex features. In terms of extracting textual representation features of scientific and technological resources, statistical models based on word frequency, topic models, and deep learning based word vector representation methods are commonly used.

TF-IDF \cite{wang2016tfidf} uses statistical methods to extract text features. The weight of a word is calculated by considering both term frequency and inverse document frequency, and the document vector representation is constructed from the weights of all words. Wu et al. proposed the TTF-LDA algorithm, which combines TF-IDF and LDA topic analysis to process academic literature abstracts \cite{wu2019ttflda}. Mikolov et al. proposed Word2Vec, which uses the CBOW and Skip-Gram models to obtain hidden-layer vector representations through word prediction tasks \cite{mikolov2013efficient}. Compared with one-hot representations, Word2Vec integrates contextual semantic information and uses distances between word vectors to represent semantic similarity.

With the rapid development of artificial intelligence, deep learning can also be used for feature extraction of scientific texts. Autoencoders can effectively learn semantic representations of text data. Eisa et al. used deep autoencoder technology to extract lexical feature sets \cite{eisa2017figureplagiarism}. Recurrent neural networks are suitable for sequence data and play an important role in text processing tasks \cite{lipton2015critical,zhao2017hinfinity}. To solve the gradient vanishing problem of long-distance dependencies, LSTM and GRU units retain long-distance semantic information through memory, forgetting, and output gates \cite{dey2017gru,zhao2017marketstate}. Encoder-decoder architectures further make it possible to map one text sequence to another and to use hidden vectors as semantic representations of scientific and technological resources.

Devlin et al. proposed the BERT pre-training model based on bidirectional Transformers, which uses multi-head self-attention to capture contextual semantics and achieves strong performance on many NLP tasks \cite{devlin2019bert}. The Transformer unit proposed by Vaswani et al. is composed of multi-head attention layers and can replace recurrent structures to obtain better parallel computing power on large corpora \cite{vaswani2017attention}. XLNet further optimizes the pre-training strategy through an autoregressive language modeling scheme \cite{yang2019xlnet}. Retrieval-oriented pre-training such as RetroMAE is closely related to scientific resource search because it improves language models for retrieval tasks \cite{xiao2022retromae}. In addition, semantic-similarity attention and hypergraph convolution can enrich scientific publication representations by incorporating high-order relations among papers, authors, keywords, and venues \cite{li2026ssahgc}. These representation learning methods provide the foundation for scientific resource acquisition, indexing, and portrait construction.

\section{Accurate Portrait of Scientific Resources}

\subsection{Scientific Resource Entity and Entity Relationship Extraction}
The construction of scientific and technological entity and concept knowledge graphs requires entity extraction and entity relationship extraction from scientific and technological resources on the network. In named entity recognition, many deep learning methods, such as convolutional neural networks and hybrid neural networks, can effectively extract scientific and technological entities from unstructured text \cite{li2017recursive,li2017variance,yadav2018survey}. Amplayo and Song proposed several network construction methods for scarce scientific literature and used full text to automatically extract entities required for network construction \cite{amplayo2016entitynetworks}. Ma and Yuan proposed a BiLSTM-CRF entity extraction method based on a feature-based named entity knowledge base to extract entities in ecological restoration technology papers \cite{ma2019bilstmcrf}. Peng and Dredze improved Chinese social media named entity recognition using jointly trained embeddings \cite{peng2015nerchinese}.

Named entity recognition also plays an important role in domain-specific texts. Zeng et al. used an LSTM-CRF model for drug named entity recognition \cite{zeng2017lstmcrfdrug}. Cao et al. combined CNN and CRF to identify entities in Chinese electronic medical records \cite{cao2019cnncrf}. Cai et al. used an LSTM-CRF model with self-attention to extract entities from Chinese electronic medical records \cite{cai2019emr}. Chen et al. proposed a semi-supervised deep learning framework for entity recognition in Chinese government documents \cite{chen2019semisupervisedgov}. Wang et al. used BERT for Chinese named entity recognition \cite{wang2019bertner}, and Cheng et al. improved Chinese short text entity linking by adding entity vectors to BERT \cite{cheng2019entitylinking}. For short scientific resource text, heterogeneous graph attention networks can be cited as a useful auxiliary method because they capture the interactions among words, documents, and topic relations \cite{hu2019hgat}.

For entity relation extraction, the main task is to extract triples of the form $\langle$entity 1, relation, entity 2$\rangle$ from unstructured text. To model entity relations, Zhang and Wang proposed replacing CNNs with RNNs to capture word dependencies \cite{zhang2015rnnrelation}. Li et al. used syntactic parse trees to recursively generate text feature representations \cite{li2015trees}. However, recurrent models may still suffer from gradient vanishing over long distances. Zhang et al. proposed a BiLSTM-based relation classification method that learns bidirectional semantic information while modeling long-distance dependencies \cite{zhang2015bilstmrelation}. Dey and Salemt proposed variants of gated recurrent units that simplify the model structure and improve relation extraction \cite{dey2017gruvariants}.

Attention mechanisms have also been introduced into relation extraction. Multi-level attention CNNs, selective attention, and sentence-level attention have been used to highlight important words, instances, or entity descriptions \cite{wang2016multilevelattention,lin2016selectiveattention,hermann2015teaching}. Attention-based methods are also used in speech recognition and multilingual relation extraction \cite{chorowski2015attention,lin2017multilingualattention,ji2017distant}. Zhou et al. proposed attention-based bidirectional LSTM networks for relation classification \cite{zhou2016attentionbilstm}, and Wang et al. proposed an LSTM semantic relation extraction method based on attention \cite{wang2018attentionlstm}. In scientific resource graphs, incomplete features and sparse structures are common; therefore, T2-GNN and federated GNN methods provide useful references for robust cross-graph node classification and graph representation \cite{huo2023t2gnn,guan2021federatedgnn}. When scientific information networks are distributed across institutions, FedSIN offers a federated self-adaptive representation learning perspective for privacy-preserving information network modeling \cite{li2026fedsin}. Federated learning with stochastic quantization is also relevant because it reduces communication costs while maintaining collaborative model learning under distributed resource settings \cite{li2022stochasticquantization}.

It is an effective method to construct the semantic representation layer of scientific and technological big data based on pre-trained models. The BERT encoder uses a bidirectional Transformer \cite{vaswani2017attentionnips}. During pre-training, Masked Language Modeling captures word-level semantics, while Next Sentence Prediction obtains sentence-level representations. Transformer introduces a self-attention mechanism to learn relationships within the source sentence, within the target sentence, and between the source and target sentences \cite{shaw2018selfattention}. The feed-forward layer uses a fully connected network and ReLU activation \cite{agarap2018relu}. Sequence-to-sequence models and structure-aware generation also provide useful insights into transforming structured data into text descriptions \cite{liu2018tabletotext}.

\subsection{Entity Extraction of Scientific and Technological Subject Words}
In scientific resource portraits, keyword extraction technology can be used to construct correlations between scientific keywords and scientific achievements. Keyword extraction extracts words or phrases that are most relevant to a text. In early information retrieval systems, keywords were often used as the retrieval basis for the entire article. Keywords still play an important role in papers, patents, and knowledge services. Scientific and technological information oriented cross-media retrieval provides a related example of integrating semantic and media information for scientific information services \cite{li2022smcr}. Keywords are also useful for text classification, clustering, and text summarization \cite{minae2021textclassification,xue2019deeplowrank,li2020graphsummarization}. By using similar keywords between texts, the convergence time of text clustering can be reduced \cite{hu2008wikipedia,sun2009knn}. Ontology-based retrieval systems further use domain concepts to improve intelligent information retrieval \cite{yang2015ontology}.

Keyword extraction can be implemented by keyword assignment or keyword extraction. Keyword assignment prepares a large keyword library and matches words in a given text, but its quality depends on the keyword library and it has limited ability to extract new words. Keyword extraction directly extracts words from the text and is more meaningful in practical applications. Word segmentation tools, such as Jieba and character-based joint segmentation models, can be used to process Chinese texts before keyword extraction \cite{ding2021jieba,wang2010segmentation}. Statistical TF-IDF and graph-based methods can then be used to rank words or phrases. YAKE extracts keywords from single documents using multiple local features \cite{campos2020yake}, while TextRank brings graph ranking ideas into keyword extraction \cite{mihalcea2004textrank}. Phrase extraction may involve word combination or generation, and phrases contain richer semantic information than isolated words.

\subsection{Relation Extraction in the Field of Scientific and Technological Achievements}
The relationship between scientific and technological achievements and subject areas is an important part of scientific resource portraits. For hierarchical subject areas, hierarchical multi-label classification can be used to associate achievements with specified subject area nodes. Hierarchical multi-label classification is a special form of multi-label classification in which labels are organized in a hierarchy, and each label may have parent labels or child labels. These associations can be used to optimize classification, but they also introduce difficulties such as data skew, complex evaluation, and hierarchy-aware decision making.

Current hierarchical multi-label classification algorithms can be divided into flat methods, local methods, global methods, and hybrid methods \cite{fall2003automated,cesa2006incremental,vens2008decision,huang2019hierarchical}. The flat method removes the association between hierarchical labels and transforms the problem into ordinary multi-label classification. Patent keyword extraction and anomaly detection methods are related to this setting because they provide feature engineering and detection tools for resource representation \cite{hu2018patent,hu2018anomaly}. The local method constructs a classifier for each hierarchical label and obtains global classification results by combining multiple classifiers. Representative local approaches include true path rule ensembles, gene ontology association, and neural-network-based local classifiers \cite{valentini2009truepath,kiritchenko2004hierarchical,cerri2011hierarchical}. The global approach builds a single classifier over the hierarchical label set \cite{silla2011survey,meng2016consensus}. Deep learning methods, such as CNNs and recursively regularized graph-CNNs, have been widely used for large-scale hierarchical text classification \cite{kim2014cnn,peng2018largehier,li2013gmphd}. Hybrid methods combine the advantages of local and global methods and often use neural networks to jointly process hierarchical information \cite{wehrmann2018hmcn,lin2009averageconsensus,fang2020identity}. Interpretable machine learning is also relevant to scientific resource portraits because knowledge service systems need transparent decisions when assigning subjects, recommending resources, and explaining entity relations \cite{li2019interpretable}.

\section{Science Resources Knowledge Service Components}
In the development and design of knowledge service systems, many studies focus on question answering, library services, smart spaces, and service componentization. Xu and Teng analyzed archives knowledge services and proposed an intelligent question answering model for multi-source archive data \cite{xu2020archiveqa}. Xu et al. designed subject knowledge services for university libraries to improve the efficiency of knowledge acquisition \cite{xu2007libraryknowledge}. Wang discussed the knowledge service ecosystem of smart libraries and designed the overall structure of a library knowledge service system \cite{wang2021smartlibrary}. Huang studied the application of big data technology in university library knowledge services \cite{huang2021bigdata}, while Ye et al. investigated smart spaces in university libraries oriented to knowledge services \cite{ye2021smartspace}. Shen and Yu constructed a system dynamics model for think tank knowledge service development \cite{shen2022thinktank}.

In terms of service components, Guo proposed a Web Service-based smart service framework that decouples functional modules through component-based logic design \cite{guo2017webservice}. Wang studied SOA service components in a user management system \cite{wang2018soa}. Distributed consensus and filtering methods provide useful technical support for distributed service infrastructure \cite{li2017kalman}. Tang and Deng designed a service component library for space launch sites, and Wang studied the description system of domain-based business components \cite{tang2014spacecomponent,wang2012domaincomponents}. Although recommendation is not the central topic of this paper, sequential recommendation and self-supervised graph co-training can be used as auxiliary references for building personalized scientific resource services, session-aware knowledge recommendation, and user behavior modeling \cite{zhou2022fmlp,xia2021graphcotrain}. Dataset distillation methods for sequential recommendation, such as Tucker-decomposition-based distillation, can further support compact service models when resources and training costs are constrained \cite{zhang2025td3}.

Based on the above studies, scientific resource knowledge service components can be divided into several layers. First, data acquisition components collect papers, patents, projects, authors, institutions, and related metadata. Second, representation components extract textual, structural, and multimodal features. Third, entity and relation extraction components construct knowledge graphs from structured and unstructured resources. Fourth, knowledge service components provide retrieval, recommendation, question answering, visualization, and decision support. Federated supervised cross-modal retrieval is also relevant to this service layer because cross-institutional scientific resources often require multimodal retrieval while preserving local data privacy \cite{li2024fedcmr}. These components jointly support accurate portraits of scientific resources.

\section{Conclusion}
With the development of big data technology, the amount of data on the Internet has grown explosively, and technological resources related to academic fields have also increased rapidly. Scientific and technological resources are mainly composed of entities closely related to academic resources, such as papers, patents, authors, institutions, and publishing units, and contain a large amount of text information. How to construct a complete and accurate representation of scientific and technological resources from structured and unstructured resources, and how to further tap the potential value of scientific and technological resources, are pressing issues.

This paper reviews accurate portraits of scientific resources and knowledge service components from several aspects, including scientific resource text representation, entity extraction, relationship extraction, keyword extraction, hierarchical subject relation extraction, and service component design. The solution is to construct accurate portraits of scientific and technological resources by combining knowledge graph technologies, deep learning, pre-training models, heterogeneous graph learning, and knowledge service design. In future work, more attention can be paid to multimodal resource modeling, explainable knowledge services, dynamic scholar portraits, and privacy-preserving scientific resource representation learning.

\begin{acks}
This work is supported by National Key R\&D Program of China (2018YFB1402600), the National Natural Science Foundation of China (61772083, 61877006, 61802028, 62002027).
\end{acks}

\bibliographystyle{yuewang_custom_unsrt}
\bibliography{references}

\end{document}